\documentclass[prl,aps,preprint,superscriptaddress]{revtex4-1}

\usepackage{graphicx}
\usepackage{amsmath}
\usepackage{dcolumn}
\usepackage{txfonts}
\usepackage{bm}

\begin{document}

\title{Tunable liquid-solid hybrid thermal metamaterials with a topology transition}

\author{Peng Jin}
\author{Jinrong Liu}
\affiliation{Department of Physics, State Key Laboratory of Surface Physics, and Key Laboratory of Micro and Nano Photonic Structures, Ministry of Education, Fudan University, Shanghai 200438, China}

\author{Liujun Xu}
\affiliation{Graduate School of China Academy of Engineering Physics, Beijing 100193, China}

\author{Jun Wang}
\affiliation{School of Physics, East China University of Science and Technology, Shanghai 200237, China}

\author{Xiaoping Ouyang}
\affiliation{School of Materials Science and Engineering, Xiangtan University, Xiangtan 411105, China}

\author{Jian-Hua Jiang}\email{Jianhuajiang@suda.edu.cn}
\affiliation{Institute of Theoretical and Applied Physics, School of Physical Science and Technology, and Collaborative Innovation Center of Suzhou Nano Science and Technology, Soochow University, Suzhou 215031, China}

\author{Jiping Huang}\email{jphuang@fudan.edu.cn}
\affiliation{Department of Physics, State Key Laboratory of Surface Physics, and Key Laboratory of Micro and Nano Photonic Structures, Ministry of Education, Fudan University, Shanghai 200438, China}

\begin{abstract}
Thermal metamaterials provide rich control of heat transport which is becoming the foundations of cutting-edge applications ranging from chip cooling to biomedical. However, due to the fundamental laws of physics, the manipulation of heat is much constrained in conventional thermal metamaterials where effective heat conduction with Onsager reciprocity dominates. Here, through the inclusion of thermal convection and breaking the Onsager reciprocity, we unveil a regime in thermal metamaterials and transformation thermotics that goes beyond effective heat conduction. By designing a liquid-solid hybrid thermal metamaterial, we demonstrate a continuous switch from thermal cloaking to thermal concentration in one device with external tuning. Underlying such a switch is a topology transition in the virtual space of the thermotic transformation which is achieved by tuning the liquid flow via external control. These discoveries illustrate the extraordinary heat transport in complex multi-component thermal metamaterials and pave the way toward an unprecedented regime of heat manipulation.
\end{abstract}

\maketitle

Developing rapidly in the past decade, thermal metamaterials~\cite{APL08,APL08-1} and transformation thermotics~\cite{SPR,PR2021,NRM2021} have greatly enriched heat manipulation which are valuable for applications ranging from thermal cloaking and camouflage~\cite{OE2012,PRL2012,PRL2014}, heat management in chips~\cite{IEEE2015,SCI2018,SCI2020,JEP2020,JEP2021,NAT2021}, daily-life energy saving~\cite{PRL2016,NATE2017,SCI2021}, to biological cell thermoregulation~\cite{CELL,PNAS1}. However, by far the development in this discipline is mainly focused on conductive thermal metamaterials~\cite{PRL2015,AM2015,APL2016,AM2018,NC2018,IJHMT2020,IJHMT2021,AM2021} where thermal transport is dominated by diffusive or effective heat conduction which is restricted by Onsager's reciprocity. The manipulation of heat in such thermal metamaterials are constrained in many aspects. Furthermore, the functions of conventional thermal metamaterials cannot be modified when the temperature settings are given~\cite{PRL2015,APL2016}, lacking the on-demand control that are desired in many situations.

Thermal convection~\cite{WIL,PRL2018,SCI2019,NM2019,NC2020,AM2020,NP2022} is another major thermal transport mechanism that has distinct nature. For a long time, thermal convection has been ignored in the study of thermal metamaterials and transformation thermotics. Only very recently, the theory of transformation thermotics has been extended to include thermal convection by developing an entirely different framework~\cite{AIPA,PRE2018,ATE2021,ATE2022}. With such theoretical advancement, a new realm is open and the development of thermal metamaterials with thermal convection is highly desirable. However, fusing thermal convection in liquids and thermal conduction in solids together to form hybrid thermal metamaterials is very challenging since these distinct thermal transport channels must balance and cooperate with each other to form stable heat and liquid flows~\cite{PNAS2,PRL2019,PRAP2020,CPB} that fulfill the underlying thermotic transformation. Technically, thermal convection and thermal conduction must be manipulated in the same location simultaneously. The design of such hybrid thermal metamaterials is thus more complicated than the conventional all-solid thermal metamaterials. To date, experimental progresses in this direction are still anticipated, although the integration of thermal convection has been exploited to achieve effective thermal conductivity (with synthetic Onsager reciprocity) in unprecedented parameter regimes~\cite{NM2019,NC2020,AM2020,PNAS3,NP2022}. Thermal metamaterials in a broad sense with simultaneous manipulation of conductive and convective heat flows beyond Onsager reciprocity is still missing.

Here, we solve this challenge by integrating thermal convection and thermal conductivity into one metamaterial-based device (a ``metadevice'') using fine-designed porous structures. Our design provides a prototype on how to control the conductive and convective properties of thermal transport independently in a single unit volume and thus opens a pathway towards a horizon for heat manipulation and applications, e.g., the continuous tunability and the attainability of a broad range of thermal transport properties. Besides, with the inclusion of thermal convection, the Onsager reciprocity breaks down and directional heat control becomes possible. Here, using the designed liquid-solid hybrid thermal metamaterials, we demonstrate experimentally the continuous switch between thermal cloaking and thermal concentration in one metadevice, which reveals the significant tunability of the hybrid thermal metamaterial. These discoveries indicate that the interplay and synergy between thermal conduction and thermal convection open a realm with unprecedented heat manipulation.

\begin{figure}[t]
\includegraphics[width=1.0\linewidth]{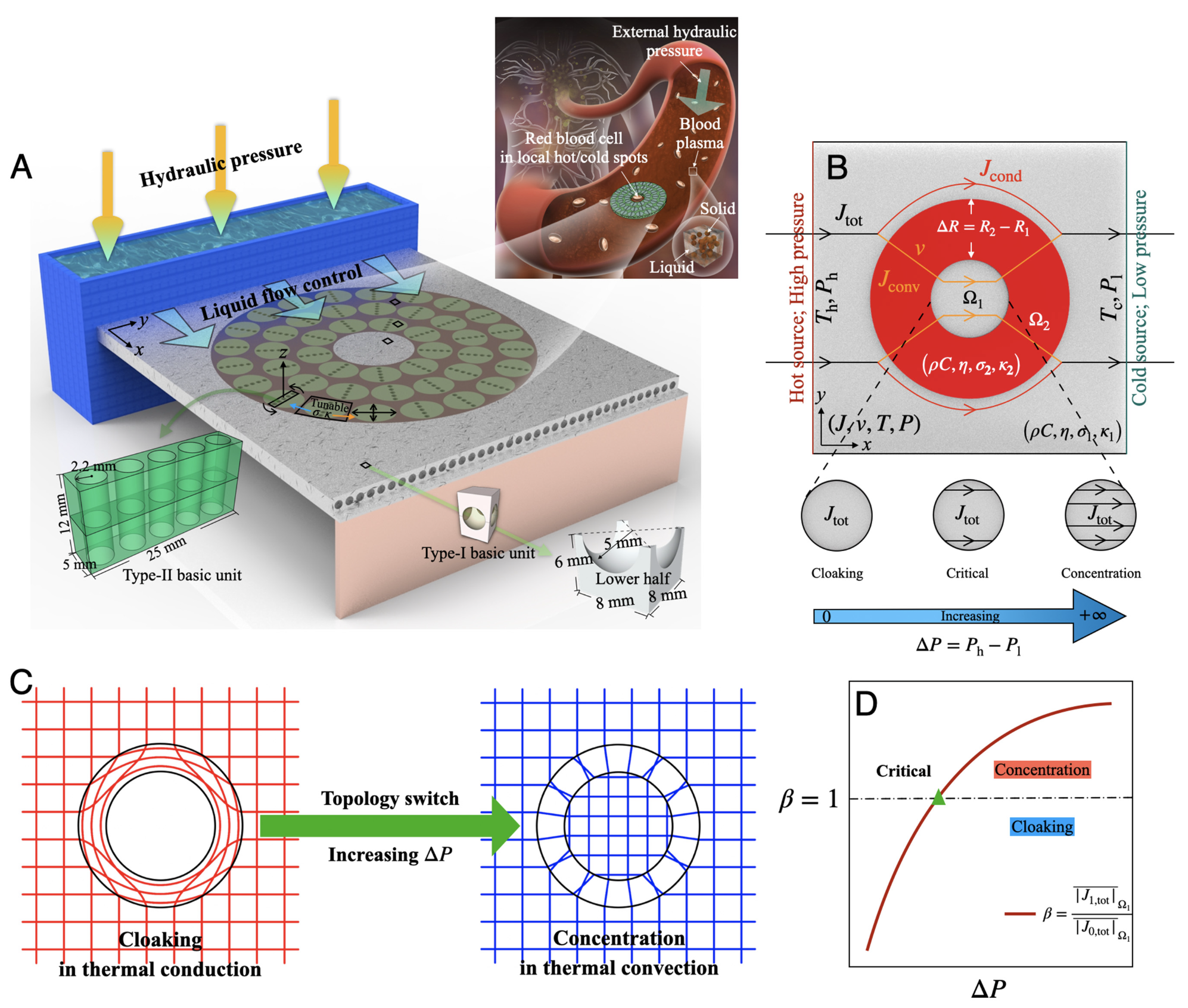}
\caption{Liquid-solid hybrid thermal metamaterial. ({\it A}) Illustration of the metadevice based on the liquid-solid hybrid thermal metamaterial. ({\it B}) Schematic depiction of the metadevice from top-view. The convective heat flux is designed to realize thermal concentration, while the conductive heat flux is designed to realize thermal cloaking. ({\it C}) The switch between thermal cloaking and thermal concentration corresponds to a topological switch in virtual space. ({\it D}) The heat flux amplification factor $\beta$ can be tuned continuously by the external hydraulic pressure. Meanwhile, the function of the metadevice is switched.}
\label{F1}
\end{figure}

\begin{figure}[t]
\includegraphics[width=1.0\linewidth]{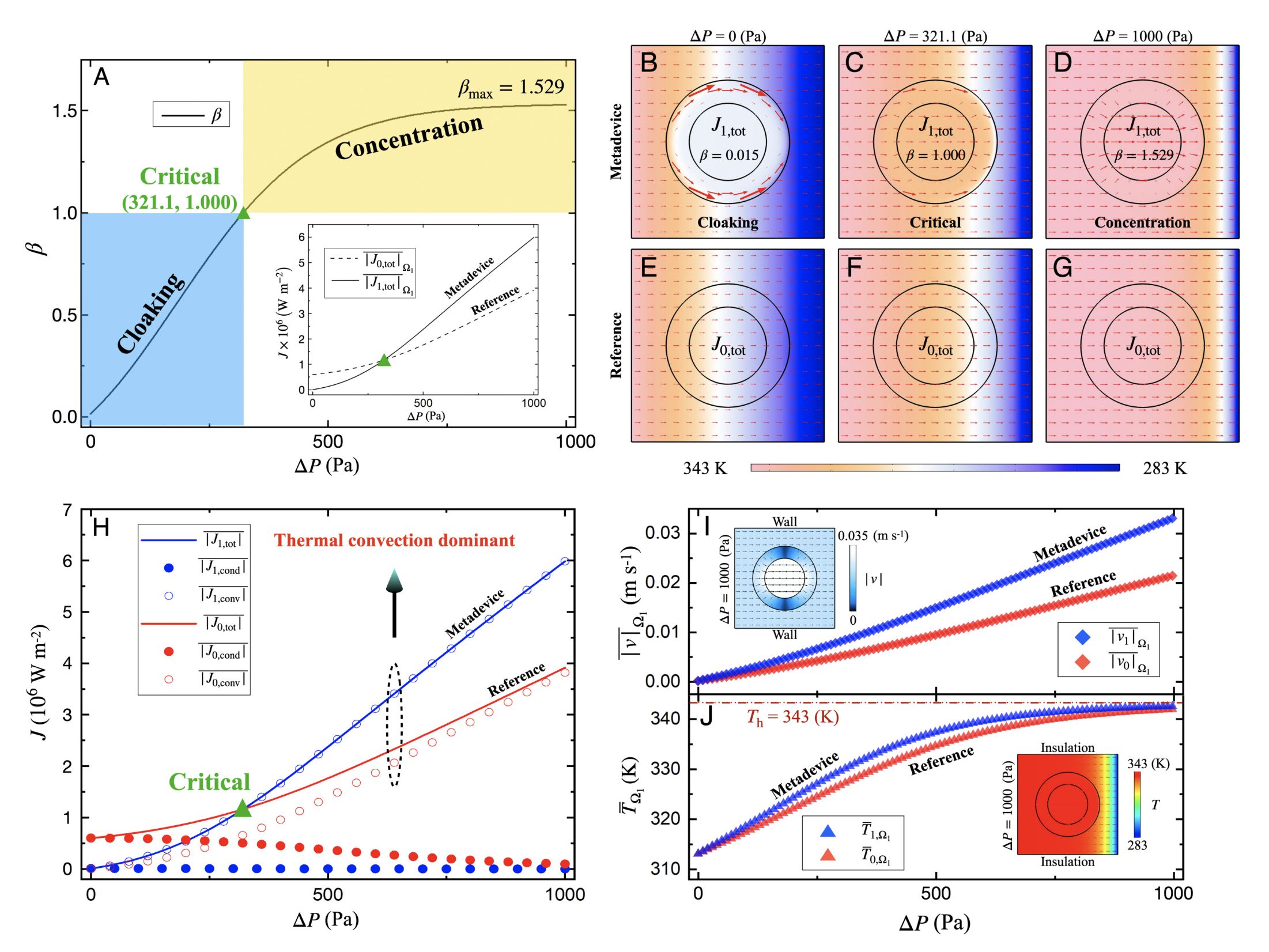}
\caption{Characterizing thermal manipulation in the metadevice via finite-element simulations. ({\it A}) Evolution of the heat flux amplification factor $\beta$ with the external hydraulic pressure difference $\Delta P$. The inset shows the dependence of the averaged amplitude of the total heat flux ($\bm{J}_{\rm tot}$) in the core region on the pressure difference $\Delta P$. The solid and dashed lines represent the heat fluxes for the metadevice (subscript 1) and the reference (subscript 0), respectively. ({\it B-D}) Temperature (color) and heat flux (vectors) profiles in the metadevice with $\Delta P$ = 0, 321.1, 1000~Pa, respectively. ({\it E-G}) Temperature (color) and heat flux (vectors) profiles in the reference with $\Delta P$ = 0, 321.1, 1000~Pa, respectively. ({\it H}) The averaged amplitude of the convective ($\bm{J}_{\rm conv}$), conductive ($\bm{J}_{\rm cond}$), and total ($\bm{J}_{\rm tot}$) heat fluxes in the core region as functions of the hydraulic pressure difference $\Delta P$. ({\it I}) and ({\it J}) Averaged fluid velocity and temperature in the core region as functions of $\Delta P$ of the metadevice and the reference, respectively. The insets show ({\it I}) the fluid velocity (color and vectors), ({\it J}) temperature (color), and heat flux (vectors) profiles in the metadevice with $\Delta P$ = 1000~Pa, respectively.}
\label{F2}
\end{figure}

\section*{Results}
\subsection*{Designing hybrid thermal metamaterial}

To create the hybrid thermal metamaterial, we use porous materials to allow thermal convection and thermal conduction to share the same space (Fig.~\ref{F1}{\it A}). The whole design involves two levels. At the first level, through the design of the basic unit, we create a porous material to achieve independent control over the thermal conduction and thermal convection properties locally. At the second level, we use the generalized theory of transformation thermotics to design the spatial profiles of the thermal conduction and thermal convection properties to achieve the targeted functions of the thermal metadevice.

When considering steady-state thermal transport in the hybrid metamaterial, the total heat flux ($\bm{J}_{\rm tot}$) is governed by the conservation equation,
\begin{equation}\label{Conservation}
\bm{\nabla}\cdot\bm{J}_{\rm tot}=\bm{\nabla}\cdot\left(\bm{J}_{\rm cond}+\bm{J}_{\rm conv}\right)=0.
\end{equation}
The total heat flux consists of the conductive heat flux described by Fourier's law $\bm{J}_{\rm cond}=-\bm{\kappa}\cdot\bm{\nabla}T$ ($\bm{\kappa}$ is the thermal conductivity tensor and $T$ is the temperature) and the convective heat flux described by Darcy's law $\bm{J}_{\rm conv}=\rho C\bm{v}T$ (velocity of the laminar fluid is $\bm{v}=-\bm{\sigma} / \eta\cdot\bm{\nabla}P$ where $\rho$, $C$, $\bm{\sigma}$, $\eta$, and $P$ are separately the mass density, the heat capacity, the permeability, the dynamic viscosity, and the hydraulic pressure of the liquid in the hybrid system). The material parameters are considered here as position-dependent. Eq.~(\ref{Conservation}) is then expressed as
\begin{equation}\label{invariant}
\bm{\nabla}\cdot\left[-\bm{\kappa}\cdot\bm{\nabla}T+\rho C\left(-\frac{\bm{\sigma}}{\eta}\cdot\bm{\nabla}P\right)T\right]=0.
\end{equation}
The above equation holds when the liquid and heat flows are stable as well as local thermal equilibrium is reached everywhere. The inclusion of thermal convection expands the physical fields to a much larger set $\left(\bm{J}_{\rm tot}, \bm{v}, T, P\right)$. The material parameter space is also extended to $\left(\bm{\kappa}, \rho, C, \bm{\sigma}, \eta\right)$. The enriched physical fields and enlarged parameter space give rise to rich thermal transport and manipulation. Here, we focus on the laminar thermal convection regime with low fluid velocity to avoid possible turbulence and nonlinear effects. Thanks to the excellent control of the hydraulic pressure and the fluid velocity, thermal transport in our liquid-solid hybrid thermal metamaterial can be tuned continuously.

The local control over the conductive and convective thermal properties is realized by the basic units design. We have two types of units. The type-I unit is a cuboid with a hemispherical region filled with water (see lower-right inset of Fig.~\ref{F1}{\it A}). Type-II unit is a cuboid with cylindrical five air holes. The effective thermal conductivity of each unit is given by $\bm{\kappa}=\left(1-\phi_l-\phi_a\right)\bm{\kappa}_{\rm s}+\phi_l\bm{\kappa}_{\rm l} + \phi_a\bm{\kappa}_{\rm a}$ where $\bm{\kappa}_{\rm s}$, $\bm{\kappa}_{\rm l}$, and $\bm{\kappa}_{\rm a}$ are separately the thermal conductivity of the solid, liquid, and air. $\phi_l$ and $\phi_a$ are the filling fraction of the liquid and air region, respectively. In each unit, the thermal conductivity can be tuned by the choice of the solid material and the filling fractions $\phi_l$ and $\phi_a$. Meanwhile, the permeability $\bm{\sigma}$ can be tuned via the geometry of the liquid or air region. For instance, in the type-II units air holes are used to deflect the liquid flow. The orientation of such units can be used to tune the permeability $\bm{\sigma}$ ({\it SI Appendix}, section S1).

\subsection*{Cotransformation of thermal conduction and convection}

The starting point of the generalized transformation thermotics theory is to note that the Eq.~(\ref{invariant}) is invariant under the coordinate transformation that satisfies the following relations ({\it SI Appendix}, section S2),
\begin{equation}
\bm{\nabla}\cdot\left(\bm{\kappa}\cdot\bm{\nabla}T\right)=0, \quad \bm{\nabla}\cdot\left(\rho C \bm{v} T\right) = 0 .
\end{equation}
In the terminology of transformation thermotics, under a transformation from real space to virtual space, the thermal conductivity transforms as follows, $\bm{\kappa}'=\bm{\Xi}\bm{\kappa}\bm{\Xi}^T/\det\bm{\Xi}$, where $\bm{\Xi}$ is the Jacobian matrix of the transformation for thermal conduction. Meanwhile, the permeability transforms according to the following relation, $\bm{\sigma}'=\bm{\Lambda}\bm{\sigma}\bm{\Lambda}^T/\det\bm{\Lambda}$ where $\bm{\Lambda}$ is the Jacobian matrix of the transformation for thermal convection. These two transformations are independent, since they are acting on different degrees of freedom.

We design a metashell structure where the thermal conductivity $\bm{\kappa}'$ distribution is targeted for thermal cloaking as dictated by our choice of the transformation $\bm{\Xi}$ (Fig.~\ref{F1}{\it A}). This transformation maps to a virtual space with a hole at the center. The hole in the virtual space is exactly the origin of the thermal cloaking effect: The heat flows cannot touch any object in the hole in the virtual space, while in real space, an object in the core region is not affected by the heat flows. On the other hand, the liquid permeability distribution $\bm{\sigma}'$ is generated by the transformation of the thermal convection $\bm{\Lambda}$, which is designed for thermal concentration. This transformation maps to a virtual space with no hole. From the geometric point of view, the virtual space with a hole is topologically distinct from the virtual spaces with no hole. Therefore, with increasing hydraulic pressure difference $\Delta P$ = $P_{\rm h}$ - $P_{\rm l}$ ($P_{\rm h}$ and $P_{\rm l}$ are the hydraulic pressure at the hot and cold sides of the metadevice, respectively), thermal convection becomes dominant and the device function switches from thermal cloaking to thermal concentration (see Fig.~\ref{F1}{\it B}). Meanwhile, the virtual space undergoes topology switch (see Fig.~\ref{F1}{\it C}). Specifically, the nontrivial topology in the virtual space for thermal cloaking implies that there are some properties robust to external conditions. These properties are the heat current in the core region. In the thermal cloaking regime, such a heat current is irrelevant with external temperature distributions. In contrast, for thermal concentration, the heat current in the core region is highly sensitive to external temperature regions. The switch between these two functions reflect the topology change in the virtual space.

To ensure the above features, a core region ($\Omega_1$) with isotropic thermal conductivity ($\kappa_1$) and permeability ($\sigma_1$) is placed at the center of the device (Fig.~\ref{F1}{\it B}). Meanwhile, outside the metadevice is the background ($\Omega_3$) with the same physical parameters as the core $\Omega_1$. In the metamaterial region ($\Omega_2$) both the thermal conductivity $\bm{\kappa}'$ and the liquid permeability $\bm{\sigma}'$ are engineered according to the transformations $\bm{\Xi}$ and $\bm{\Lambda}$. Here, we focus on the two-dimensional transformation described in the cylindrical coordinate $(r\cos\theta, r\sin\theta)$. In the device region, the both the thermal conductivity $\bm{\kappa}'$ and the liquid permeability $\bm{\sigma}'$ are anisotropic. They are expressed as the diagonal tensors,  $\bm{\kappa_2}$ = diag($\kappa_{rr}$,$\kappa_{\theta\theta}$) with $\kappa_{rr}\kappa_{\theta\theta}=\kappa_1^2$ and $\bm{\sigma_2}$ = diag($\sigma_{rr}$,$\sigma_{\theta\theta}$) with $\sigma_{rr}\sigma_{\theta\theta}=\sigma_1^2$. Here, we choose transformations with weak position dependence in the region ($\Omega_2$) and realize them with approximately $r$-independent structures. The specific forms of these quantities are given in {\it SI Appendix}, section S2.

To characterize quantitatively the function of the designed metadevice, we introduce the heat flux amplification factor $\beta$ which is given by the averaged amplitude of the total heat flux in the core region ($\Omega_1$) over the same quantity when the system is as uniform as the background (henceforth denoted as ``the reference''). The $\beta$ factor characterizes the function of the metadevice,
\begin{equation}
\beta\left(\Delta P\right)=\frac{\overline{{|\bm{J}_{1,\rm tot}\left(\Delta P\right)|}}_{\Omega_1}}{\overline{{|\bm{J}_{0,\rm tot}\left(\Delta P\right)|}}_{\Omega_1}}
\begin{cases}\beta~>~1,~\rm{concentration,}\\
\beta~=~1,~\rm{critical~point,}\\
0~<~\beta~<~1,~\rm{cloaking,}
\end{cases}
\end{equation}
where $\bm{J}_{1,\rm tot}$ and $\bm{J}_{0,\rm tot}$ are, respectively, the total heat flux in the core region for the metadevice and the reference. Remarkably, the different functions listed above can be achieved in the designed metadevice by tuning the external hydraulic pressure $\Delta P$ (Fig.~\ref{F1}{\it D}).

\subsection*{Simulation and characterization}

Before going into the experiments, we first perform simulations based on the designed distributions of the effective parameters ({\it SI Appendix}, section S3). The simulation box is 0.1 $\times$ 0.1~mm$^{2}$ with $R_1=0.02$~mm and $R_2=0.032$~mm. We choose the following parameters: The homogeneous background and the core region have the same parameters: the isotropic thermal conductivity $\kappa_1 = 1$~W~m$^{-1}$~K$^{-1}$ and the isotropic liquid permeability $\sigma_1 = 10^{-12}$~m$^{2}$. In the metamaterial region ($\Omega_2$), $\bm{\kappa}_2$ = diag(0.1,10)~W~m$^{-1}$~K$^{-1}$ and $\bm{\sigma}_2$ = diag(10,0.1) $\times$ $10^{-12}$~m$^{2}$. The left (right) terminal of the system is connected with a hot (cold) source of temperature $T_{\rm h}$ = 343~K ($T_{\rm c}$ = 283~K). The averaged amplitude of the total heat flux ($\bm{J}_{\rm tot}$) in the core region ($\Omega_1$) is calculated from finite-element simulations when the external hydraulic pressure $\Delta P$ is increased from 0 to 1000~Pa with an interval of 10~Pa. As shown in Fig.~\ref{F2}{\it A}, the heat flux amplification factor $\beta$ indeed increases with the hydraulic pressure difference $\Delta P$.

Figs.~\ref{F2}{\it B-D} show from finite-element simulations how the functions of the metadevice can be tuned by the external hydraulic pressure. In these figures, the distributions of the temperature (represented by the color profiles) and the total heat flux (depicted by the red arrows) are presented. For $\Delta P$ = 0, the $\beta$ factor is minimal $\beta$ = 0.015, indicating that the core region $\Omega_1$ is cloaked from the heat flows. Indeed, the heat flux is deflected by the metamaterial region, as shown in Fig.~\ref{F2}{\it B}. For $\Delta P$ = 321.1~Pa (Fig.~\ref{F2}{\it C}), $\beta$ = 1 and the heat flux in the core region is the same as the heat flux in the background, indicating the critical situation. For a larger $\Delta P$ = 1000~Pa (Fig.~\ref{F2}{\it D}), the $\beta$ factor is more significant than one, leading to heat flux focused on the core region, i.e., thermal concentration. The above tuning of thermal transport is achieved without disturbing the temperature field in the background, demonstrating an unprecedented regime in heat manipulation.

For comparison, we give the distributions of the heat flux and temperature in Figs.~\ref{F2}{\it E-G} for the reference system under the same external hydraulic pressure and temperature setup with Figs.~\ref{F2}{\it B-D}. In the background region ($\Omega_3$), our device has nearly the same heat flux and temperature distributions as those in the reference system. In contrast, the heat flux and temperature profiles are substantially manipulated in the metamaterial region $\Omega_2$ and in the core region $\Omega_1$. These results demonstrate the continuous tunability and the power of the hybrid thermal metamaterial in the manipulation of heat.

When $\Delta P$ is large, thermal convection is dominant (see Fig.~\ref{F2}{\it H}) which leads to significant thermal concentration effect. In this regime, the isotherm line is pushed close to the cold terminal, and the temperature of the core region approaches $T_{\rm h}$. In this limit, $\beta$ reaches its maximum value
\begin{equation}\label{max}
\beta_{\rm max}=\left(\frac{R_2}{R_1}\right)^{1-\frac{\sigma_1}{\sigma_{rr}}} 
\end{equation}
which depends on the geometric and material parameters of the metadevice ({\it SI Appendix}, section S4). As shown in Fig.~\ref{F2}{\it A}, the $\beta$ factor can be tuned continuously via the external hydraulic pressure from 0 to a maximum value of $\beta_{\rm max}$ = 1.529, which agrees with Eq.~(\ref{max}). The dependences of the maximum heat amplification factor $\beta_{\rm max}$ on both the geometric and material parameters of the hybrid metamaterial are discussed in more details in {\it SI Appendix}, section S4.

Figs.~\ref{F2}{\it I} and {\it J} present the averaged liquid velocity and temperature in the core region as functions of the external hydraulic pressure, respectively. We note that the hybrid metamaterial significantly modifies the averaged fluid velocity, demonstrating manipulation of liquid flows in parallel with heat manipulation. The averaged temperature in the core region is also considerably changed. The maximum temperature difference between the averaged temperatures of the metadevice and the reference can reach $3^{\circ}C$. With increasing hydraulic pressure difference $\Delta P$, the averaged temperature of the core region approaches $T_{\rm h}$ = 343~K, which is consistent with Figs.~\ref{F2}{\it D} and {\it G}. In particular, the insets in Figs.~\ref{F2}{\it I} and {\it J} show the liquid velocity and temperature profiles when $\Delta P$ = 1000~Pa, demonstrating a concrete thermal concentration effect.

For the completeness of the investigations, we also design and study another liquid-solid hybrid thermal metamaterial where convective thermal transport leads to thermal cloaking and conductive thermal transport leads to thermal concentration in {\it SI Appendix}, section S5. We find that with the increase of the external hydraulic pressure, thermal transport switches from conduction dominant to convection dominant. Meanwhile, the function of the thermal metamaterial switches from thermal concentration to thermal cloaking, leading to a topological switch in the virtual space from trivial to topological.

\subsection*{Experimental realization and measurements}

\begin{figure}[t]
\includegraphics[width=1.0\linewidth]{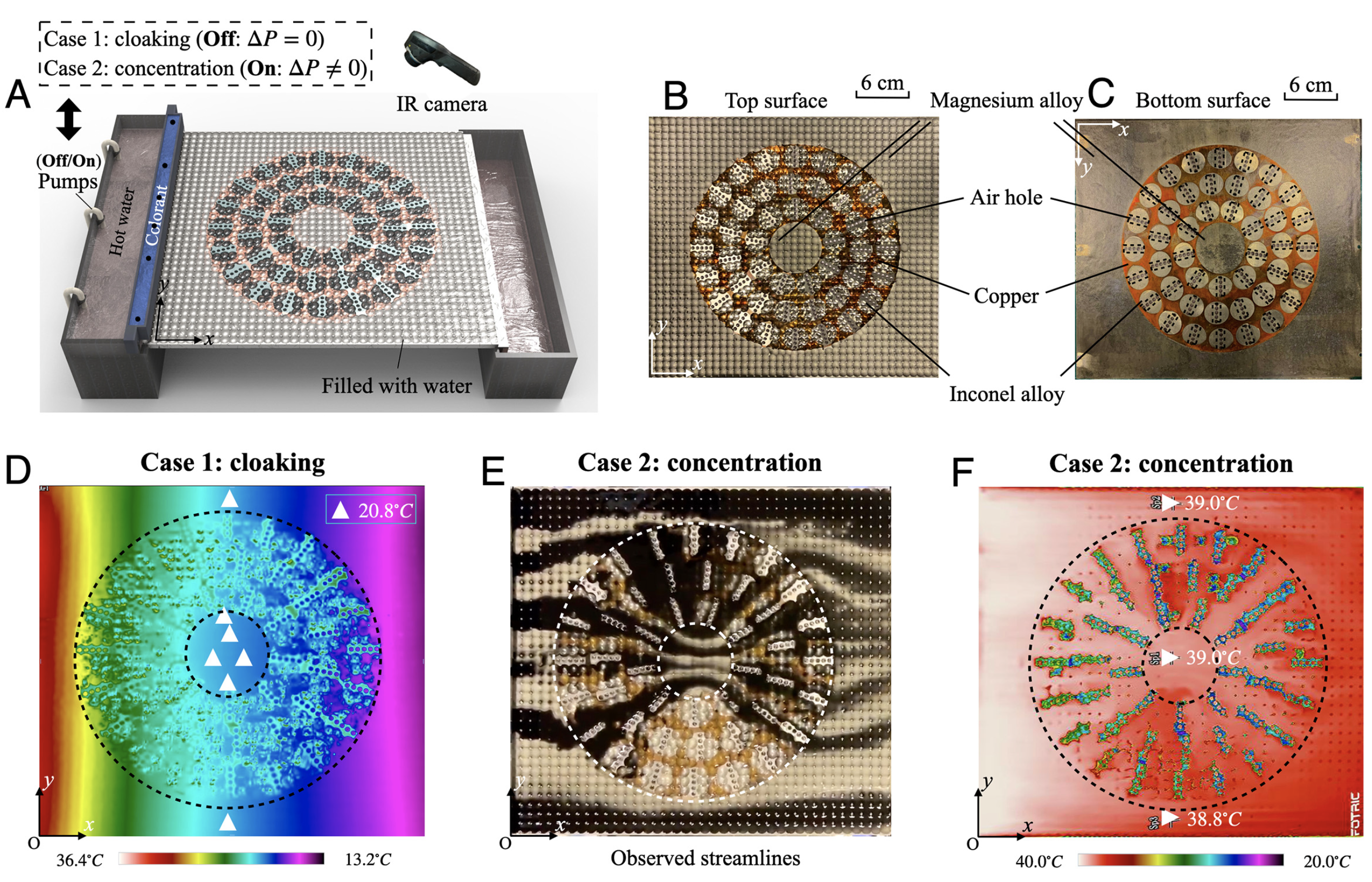}
\caption{Experimental setup and measurements. ({\it A}) Schematic of the experimental setup with different boundary conditions: pumps off, $\Delta P$ = 0; pumps on, $\Delta P$ $\ne$ 0. ({\it B}) and ({\it C}) Photos of the top and the bottom of the sample. Scale bar is 6~cm. ({\it D}) Measured temperature profile of the thermal metadevice at $\Delta P$ = 0. White triangles denote the positions with the temperature of 20.8~$^{\circ}C$. ({\it E}) Observed streamlines of the thermal metadevice at $\Delta P$ $\ne$ 0. ({\it F}) Measured temperature profile of the thermal metadevice at $\Delta P$ $\ne$ 0. Horizontal white triangles denote the positions with the temperature (from up to down) of 39.0~$^{\circ}C$, 39.0~$^{\circ}C$, and 38.8~$^{\circ}C$, respectively.}
\label{F3}
\end{figure}

The metadevice with the liquid-solid hybrid thermal metamaterial is shown in Fig.~\ref{F3}{\it A}. In the background region $\Omega_3$ and the core region $\Omega_1$, the metadevice is based on type-I units made of magnesium alloy. In the shell-like metamaterial region $\Omega_2$, both type-I and type-II units are used. Note that in the shining elliptic regions, inconel alloys are used in place of the copper to tune the thermal conductivity, particularly by adjusting the ellipses' major and minor axes. The type-II units are in the middle of the elliptic regions that can tune the permeability $\sigma$.

The liquid-solid hybrid metamaterial is mostly made of the type-I unit through which water can flow. The water is connected to a hot tank and a cold tank on the two sides, which serve as heat baths. We exploit three water pumps to tune the hydraulic pressures. As shown on the left side of Fig.~\ref{F3}{\it A}, there are two types of boundary conditions adopted in our experiments. Boundary condition I keeps no water pump on, hence $\Delta P$ = 0. Boundary condition II keeps three water pumps on, each with a flow of 200~mL~min$^{-1}$. These water pumps drive water flow filling up the hot tank and going from the hot source to the cold drain through the metadevice, which provides steady and controllable hydraulic pressure. With such designs, the water-filled background material has effective thermal conductivity with $\kappa_1$ = 27.2~W~m$^{-1}$~K$^{-1}$ and liquid permeability with $\sigma_1$ = 2.68 $\times$ $10^{-9}$~m$^{2}$. The fabricated hybrid metamaterial realizes the effective thermal conductivity $\bm{\kappa}_2$ = diag(16.6, 38.8)~W~m$^{-1}$~K$^{-1}$ and liquid permeability $\bm{\sigma}_2$ = diag(2.26, 1.02) $\times$ $10^{-9}$~m$^{2}$ in the region $\Omega_2$. See experimental section for the sample's size and effective physical characteristics.

We now demonstrate the switch between thermal cloaking and thermal concentration via controlled hydrodynamics. Under boundary condition I, $\Delta P$ = 0, we measure the temperature profile in the metadevice. As shown in Fig.~\ref{F3}{\it D}, the temperature distribution recorded by the infrared camera offers a perfect pattern of thermal cloaking that is consistent with the simulation results in Fig.~\ref{F2}{\it B}. In particular, the core region has a pretty uniform temperature distribution around $20.8^\circ C$ which indicates no conductive heat flow in the core region. Moreover, the temperature profile in the background region is nearly undisturbed. Therefore, it can be explained that $\beta$ < 1, and the metadevice presents the cloaking function.

Boundary condition II is used for the implementation of the thermal concentration. We ensure that thermal convection is dominant in such cases (see Experimental Section). Guided by the finite-element simulation results in the insets of Figs.~\ref{F2}{\it I} and {\it J}, we experimentally demonstrate the performance of thermal concentration from the perspective of fluid flow and temperature profile. To visualize the fluid flow, six holes (black dots marked in Fig.~\ref{F3}{\it A}) are punched under the container of a colorant (consisting alkanes and toner). When the system reaches a nonequilibrium steady-state, the colorant is dripped on the left boundary of the metadevice almost simultaneously and equidistantly through the six holes. Fig.~\ref{F3}{\it E} displays the six streamlines. The four streamlines in the middle are concentrated into the core region, and all the streamlines outside the region $\Omega_2$ are only slightly distorted. These results indicate that the core region has a larger flow with the same cross-sectional area (or, say, larger fluid velocity) than the background region. Note that in the $\Omega_2$ region, the colorant distributions in the upper and lower parts are different. The underlying reason is that the distribution of type-II basic units is asymmetric, and the six holes are asymmetric with respect to the upper and lower parts as well. Nevertheless, the concentration of the fluid flow is clearly visible in Fig.~\ref{F3}{\it E}. We then measure the temperature profile in the metadevice under the same condition. The measured temperature profile (see Fig.~\ref{F3}{\it F}) exhibits several features consistent with the simulation results in Fig.~\ref{F2}{\it D}. First, the overall temperature of the metadevice is higher than in the cloaking case. Moreover, the temperature gradient is pushed to the right side of the metadevice. There are visible correlations between the temperature profile and liquid flow profile, indicating that the thermal transport is now dominated by the convective heat flow carried by the water. The convective heat flow in the core region is larger than that in the background region because of $v_{\Omega_1} > v_{\Omega_3}$ with $T_{\Omega_1}$ $\approx$ $T_{\Omega_3}$ (see the white triangles in Fig.~\ref{F3}{\it F}). At this time, $\beta$ > 1. The above results are consistent with the simulated results in the insets of Figs.~\ref{F2}{\it I} and {\it J} (see {\it SI Appendix}, section S6 for more simulation results). Therefore, the metadevice is tuned into thermal concentration by increasing the external hydraulic pressure.

\section*{Conclusion and discussion}

We propose an approach to realize continuously tunable liquid-solid hybrid thermal metamaterial based on cotransformation of thermal conduction and thermal convection. With such an approach, we realize in experiments a metadevice based on a liquid-solid hybrid thermal metamaterial, which can achieve a continuous switch between thermal cloaking and thermal concentration via controlled hydrodynamics. Such a switch corresponds to a topology transition in the virtual space. A salient feature of our liquid-solid hybrid thermal metadevice is that the heat flow in the central region of the device can be continuously tuned from near zero to a tremendous value. Such significant tuning is achieved without disturbing the temperature field in the background, demonstrating extraordinary heat manipulation that cannot be achieved in conventional thermal metamaterials. Liquid-solid hybrid thermal metamaterials can be valuable in various applications such as thermal illusion and camouflage (see {\it SI Appendix}, section S7 for more discussions), cooling and heat management in electronic devices, sustainable infrastructures and intelligent heat control in smart materials and machines.

Now we discuss the topology transition of the proposed hybrid metamaterial. From a physical point of view, the transformation theory is a bridge linking geometric and (material) parametric transformations, so the topological feature of a geometric transformation is equivalently reflected in parameters. Consequently, we can discuss topology through (effective) thermal conductivity. For example, the effective thermal conductivity of a core-shell structure (for thermal cloaking) can be referred to as a topological invariant, which does not change with the thermal conductivity of the core. In other words, a topologically-nontrivial geometric transformation, to some extent, gives rise to singular (zero or infinite) thermal conductivity, which constructs an isolated region from the environment. Therefore, transformation thermotics also plays a crucial role in relating geometric topology with parametric topology. In general, zero or infinite thermal conductivity is topologically nontrivial (This is in analog with the fact that zero-index metamaterials are topologically nontrivial: they enable reflectionless transport in real-space and are equivalent to materials with Dirac dispersions in wave-vector space).

In view of the existing metamaterial research about multifunctional metadevices~\cite{APL,APL2016,SP,PA,Cry,IJHMT,EPL}, we find a notable difference between our study and those works: In those works, the switch of functions are achieved by changing the structural parameters of the metadevices. In contrast, in our study, we do not need to change the structure. Instead, the switch of function between thermal cloaking and thermal concentration is achieved by tuning the external hydraulic pressure. It is remarkable that here hydrodynamics drives the switch between nontrivial topology in the virtual space for thermal cloaking and trivial topology in the virtual space for thermal concentration. In addition, from the perspective of function switching, we utilize fluid flow to achieve switching between different functions, remarkably, continuous tuning between functions. The new mechanism unveiled here is expected to bring about unprecedented applications. Our method provides a reference for the continuous tunability of intensity in other physical domains, including electromagnetics, acoustics, mechanics, water flow, and particle dynamics.

Last but not least, we envision a potential application of our metadevice for the regulation of local hot/cold spots in biological cells or tissues without disturbing the human body's environment. The extracellular fluid in the organism is a mixture of liquids and solid-like components (Fig.~\ref{F1}{\it A}). We may model it as a porous medium and design liquid-solid hybrid metamaterials on a specific scale. Assume that in a uniform environment, we can always find the direction of the local heat flux. Based on the transformation theory, our structure is axisymmetric and causes no environmental deflection. Therefore, our model can be considered as placed along the direction of the local heat flux. The adiabatic boundaries at the top and bottom are equivalent to open boundaries, as there is zero heat flux in the vertical direction. In other words, our model is a useful abstraction of the actual environment. Driven by external hydraulic pressure, we can realize the regulation of local hot/cold spots in living cells or biological tissues based on the designed metamaterials. Here, local hot/cold spots usually associate with transient bias temperature. By controlling the hydraulic pressure, the bias distribution will disappear more quickly. For the local hot/cold spots, an increase in fluid velocity and heat flux can accelerate the transition to the suitable temperature near the steady state. Besides, in the biological environment, the extra flux will promote the balance of chemical concentrations like ATPs and CO$_2$, which help the recovery of biological functions. Meanwhile, it’s also feasible to implant such microscopic metamaterials in the human body for medical care. Our design works without disturbing the background’s thermal/fluid environment, which is vital enough for human health.

\section*{Experimental section}

A prepared annular metashell with radii of $R_1$ = 30, $R_2$ = 120~mm and thickness of 6~mm (Copper with $\kappa$ = 400~W~m$^{-1}$~K$^{-1}$ inlaid with Inconel 625 alloy with $\kappa$ = 9.8~W~m$^{-1}$~K$^{-1}$) is embedded in the center of a pure background with the size of 328 $\times$ 300 $\times$ 6~mm$^{3}$ (Magnesium alloy AZ91D with $\kappa$ = 72.7~W~m$^{-1}$~K$^{-1}$), forming a fabricated thermal metadevice with two sides, as shown in Figs.~\ref{F3}{\it B} and {\it C}. The metashell is approximately divided into three layers of basic units (size: 30.5 $\times$ 30 $\times$ 6~mm$^{3}$; quantities from inside out: 9, 15, and 20) and elliptically shaped Inconel alloys (major axis: 30~mm, minor axis: 26~mm) are embedded in Copper in each of basic unit. The entire metadevice outside the black dashed boxes (length: 25~mm, width: 5~mm) in Fig.~\ref{F3}{\it C} is divided into a set of type-I units with a size of 8 $\times$ 8 $\times$ 6~mm$^{3}$, and each type-I unit has an inner central tunneling hemisphere with a radius of $R$ = 5~mm. Therefore, each type-I unit's porosity $\phi_l$ is calculated as 0.6309. Using the volume-averaging method, the effective thermal conductivity of water-filled ($\kappa_l$ = 0.6~W~m$^{-1}$~K$^{-1}$) Magnesium alloy, Copper, and Inconel alloy are calculated as 27.2, 148.0, and 4.0~W~m$^{-1}$~K$^{-1}$, respectively. To ensure uniform thermal conductivity of each elliptical Inconel alloy, air holes ($R$ = 2.2~mm, porosity $\phi_a$ = 0.60, $\kappa$ = 0.026~W~m$^{-1}$~K$^{-1}$) are drilled inside each black dashed box (type-II unit), leading to its effective thermal conductivity is the same as that of the type-I unit in each basic unit. In our metadevice, the ratio between the current air hole size (diameter: 4.4~mm) and sample size (scale: 328~mm) is about 0.013, which is small enough to be consistent with the presented theory. From accurate finite-element simulations in {\it SI Appendix}, section S1, the effective thermal conductivity of the metashell is diag(16.6,38.8)~W~m$^{-1}$~K$^{-1}$.

The designed metadevice contains two types of units, which have different effective permeability. The type-II units, as baffles, are used to deflect the liquid flow. Their permeability is set as 10$^{-18}$~m$^{2}$. To calculate the effective permeability of the type-I units, we consider an estimation method based on experimental boundary conditions. The unidirectional steady flow in the porous material is described by Darcy's law as follows:
\begin{equation}\label{flow}
Q = \frac{\sigma A}{\eta L}\Delta P,
\end{equation}
where $\sigma$ is the permeability and $\eta$ = 0.001~Pa·s is the dynamic viscosity. $L$ = 0.3~m and $\Delta P$ = 1000~Pa are the scale and the external hydraulic pressure between the two ends of system, respectively. $A$ is cross-sectional area of the flow. $Q$ is water flow pumped into the metadevice. In the experiment, we use three peristaltic pumps (inner diameter: $D$ = 3~mm, outer diameter: $D$ = 6~mm) to pump water flow into the tank (each pump: 200~mL~min$^{-1}$). When the liquid level overflows the sink, water flows from the boundary into the metadevice. Thus, the total water flow $Q$ is 1 $\times$ $10^{-5}$~m$^{3}$~s$^{-1}$. The experimental effective cross-sectional area ($A_e$) can be calculated as
\begin{equation}
A_e = \frac{N V_{\rm Unit}\phi}{L_{\rm Unit}},
\end{equation}
where $V_{\rm Unit}$ = 8 $\times$ 8 $\times$ 6~mm$^{3}$ and $L_{\rm Unit}$ = 8~mm are the volume of the type-I unit and the length of the type-I unit along flow direction. $N$ = 37 is the number of type-I unit covered by flow along the direction of the cross-section. Finally, $A_e$ is calculated as 1.12 $\times$ $10^{-3}$~m$^{2}$. From Eq.~(\ref{flow}), the effective permeability of the type-I units is calculated as 2.68 $\times$ $10^{-9}$~m$^{2}$.

From accurate finite-element simulations ({\it SI Appendix}, section S1), the effective permeability of the metashell is diag(2.26,1.02) $\times$ $10^{-9}$~m$^{2}$.

In the finite-element simulations, $\sigma$, $\eta$, $L$, and $\Delta P$ are set as $10^{-12}$~m$^{2}$, $10^{-3}$~Pa·s, $10^{-4}$~m, and 1000~Pa. Therefore, the fluid velocity is 0.01~m~s$^{-1}$, which is approximately equal to the experimental fluid velocity $v=Q/A_e$.

The Reynolds number is calculated as
\begin{equation}
\rm{Re} = \frac{\rho v d}{\mu},
\end{equation}
where $\rho$ = 1000~kg~m$^{-3}$ is the density of the water, $v$ = 0.009~m~s$^{-1}$ is the fluid velocity, $d$ = 0.005~m is the height of liquid level in metadevice, and $\mu$ = 0.001~Pa·s is the dynamic viscosity. Finally, $\rm{Re}$ = 45, which satisfies the condition of steady flow. 

Therefore, flow boundaries in the experiment ensure that thermal convection is dominant in the heat transport.

At the left and right sides of the metadevice, a heat exchange process between the liquid and the solid leads to a slight mismatch between the fixed temperature boundary in the simulation and the heat exchange boundary in the experiment. For the general situation, the hot and cold source will be much larger than the metadevice. Thus the fixed temperature boundary is more suitable. In our experiment, since the left hot side is with great thermal contact, the difference between fixed and heat exchange boundaries can be neglected. On the right cold side, the difference is mainly equivalent to a slight change of temperature of the cold boundary $T_{\rm c}$, which slightly promotes the pushed-right temperature distribution in Fig.~\ref{F2}{\it D}. See the analyzation in {\it SI Appendix}, section S8. As the concentration of liquid flow (mainly contributing to the total heat flux) and the relationship between the temperature of the core region and the background region are not affected, the heat flux amplification factor $\beta$ of the experiment is almost not affected. Our metadevice has successfully realized the desirable effect. Besides, the above issue can be reduced by promoting thermal contact of the cold side, such as increasing the contact area, stirring the cold water with pumps, and using self-adaptive thermoelectrical coolers as cold sources~\cite{AM2022}.

To prevent heat loss in the air, we reduce the air convection in the laboratory (closing doors, windows, and air conditioning), and ensure that the average temperature of the experimental sample is consistent with the room temperature. The real experimental setup is shown in {\it SI Appendix}, Fig.~S9.

\subsection*{Data Availability}
All study data are included in the article and/or SI Appendix.

\section*{Acknowledgments}
J.H. acknowledges financial support from the National Natural Science Foundation of China under Grants No.~11725521 and No.~12035004 and from the Science and Technology Commission of Shanghai Municipality under Grant No.~20JC1414700. J.-H.J. acknowledges financial support from the National Natural Science Foundation of China under Grants No.~12125504 and No.~12074281.


\clearpage
\newpage
\begin{thebibliography}{99}
\bibliographystyle{apsrev4-1}

\bibitem{APL08} C. Fan, Y. Gao, J. Huang, Shaped graded materials with an apparent negative thermal conductivity. {\it Appl.\ Phys.\ Lett.\/} {\bf 92}, 251907 (2008).

\bibitem{APL08-1} T. Chen, C.-N. Weng, J.-S. Chen, Cloak for curvilinearly anisotropic media in conduction. {\it Appl.\ Phys.\ Lett.\/} {\bf 93}, 114103 (2008).

\bibitem{SPR} J. Huang, {\it Theoretical Thermotics: Transformation Thermotics and Extended Theories for Thermal Metamaterials\/} (Springer Nature, 2020).

\bibitem{PR2021} S. Yang, J. Wang, G. Dai, F. Yang, J. Huang, Controlling macroscopic heat transfer with thermal metamaterials: Theory, experiment and application. {\it Phys.\ Rep.\/} {\bf 908}, 1-65 (2021).

\bibitem{NRM2021} Y. Li {\it et al.}, Transforming heat transfer with thermal metamaterials and devices. {\it Nat.\ Rev.\ Mater.\/} {\bf 6}, 488-507 (2021).

\bibitem{OE2012} S. Guenneau, C. Amra, D. Veynante, Transformation thermodynamics: cloaking and concentrating heat flux. {\it Opt.\ Express\/} {\bf 20}, 8207-8218 (2012).

\bibitem{PRL2012} S. Narayana, Y. Sato, Heat Flux Manipulation with Engineered Thermal Materials. {\it Phys.\ Rev.\ Lett.\/} {\bf 108}, 214303 (2012).

\bibitem{PRL2014} T. Han {\it et al.}, Experimental Demonstration of a Bilayer Thermal Cloak. {\it Phys.\ Rev.\ Lett.\/} {\bf 112}, 054302 (2014).

\bibitem{IEEE2015} E. M. Dede, P. Schmalenberg, T. Nomura, M. Ishigaki, Design of Anisotropic Thermal Conductivity in Multilayer Printed Circuit Boards. {\it IEEE\ Trans.\ Compon.\ Packag.\ Manuf.\ Technol.\/} {\bf 5}, 1763–1774 (2015).

\bibitem{SCI2018} S. Li {\it et al.}, High thermal conductivity in cubicboron arsenide crystals. {\it Science\/} {\bf 361}, 579–581 (2018).

\bibitem{SCI2020} K. Chen {\it et al.}, Ultrahigh thermal conductivity in isotope-enriched cubic boron nitride. {\it Science\/} {\bf 367}, 555–559 (2020).

\bibitem{JEP2020} K. W. Jung {\it et al.}, Thermal and Manufacturing Design Considerations for Silicon-Based Embedded Microchannel-3D Manifold Coolers (EMMCs): Part 1-Experimental Study of Single-Phase Cooling Performance With R-245fa. {\it J.\ Electron.\ Packag.\/} {\bf 142}, 031117 (2020).

\bibitem{JEP2021} J. C. Kim {\it et al.}, Recent Advances in Thermal Metamaterials and Their Future Applications for Electronics Packaging. {\it J.\ Electron.\ Packag.\/} {\bf 143}, 010801 (2021).

\bibitem{NAT2021} S. E. Kim {\it et al.}, Extremely anisotropic van der Waals thermal conductors. {\it Nature\/} {\bf 597}, 660–665 (2021).

\bibitem{PRL2016} X. Shen, Y. Li, C. Jiang, J. Huang, Temperature Trapping: Energy-Free Maintenance of Constant Temperatures as Ambient Temperature Gradients Change. {\it Phys.\ Rev.\ Lett.\/} {\bf 117}, 055501 (2016).

\bibitem{NATE2017} E. A. Goldstein, A. P. Raman, S. Fan, Sub-ambient non-evaporative fluid cooling with the sky. {\it Nat.\ Energy\/} {\bf 2}, 17143 (2017).

\bibitem{SCI2021} K. Tang {\it et al.}, Temperature-adaptive radiative coating for all-season household thermal regulation. {\it Science\/} {\bf 374}, 1504–1509 (2021).

\bibitem{CELL} L. B. Persson, V. S. Ambati, O. Brandman, Cellular control of viscosity counters changes in temperature and energy availability. {\it Cell\/}, {\bf 183}, 1572-1585 (2020).

\bibitem{PNAS1} K. Oyama {\it et al.}, Heat-hypersensitive mutants of ryanodine receptor type 1 revealed by microscopic heating. {\it Proc.\ Natl.\ Acad.\ Sci.\ U.S.A.\/} {\bf 119}, e2201286119 (2022).

\bibitem{PRL2015} Y. Li {\it et al.}, Temperature-dependent transformation thermotics: From switchable thermal cloaks to macroscopic thermal diodes. {\it Phys.\ Rev.\ Lett.\/} {\bf 115}, 195503 (2015).

\bibitem{AM2015} T. Yang {\it et al.}, Invisible Sensors: Simultaneous Sensing and Camouflaging in Multiphysical Fields. {\it Adv.\ Mater.\/} {\bf 27}, 7752–7758 (2015).

\bibitem{APL2016} X. Shen, Y. Li, C. Jiang, Y. Ni, J. Huang, Thermal cloak-concentrator. {\it Appl.\ Phys.\ Lett.\/} {\bf 109}, 031907 (2016).

\bibitem{AM2018} T. Han {\it et al.}, Full-Parameter Omnidirectional Thermal Metadevices of Anisotropic Geometry. {\it Adv.\ Mater.\/} {\bf 30}, 1804019 (2018).

\bibitem{NC2018} Y. Li, X. Bai, T. Yang, H. Luo, C.-W. Qiu, Structured thermal surface for radiative camouflage. {\it Nat.\ Commun.\/} {\bf 9}, 273 (2018).

\bibitem{IJHMT2020} P. Jin, L. Xu, T. Jiang, L. Zhang, J. Huang, Making thermal sensors accurate and invisible with an anisotropic monolayer scheme. {\it Int.\ J.\ Heat\ Mass\ Transf.\/} {\bf 163}, 120437 (2020).

\bibitem{IJHMT2021} P. Jin {\it et al.}, Particle swarm optimization for realizing bilayer thermal sensors with bulk isotropic materials. {\it Int.\ J.\ Heat\ Mass\ Transf.\/} {\bf 172}, 121177 (2021).

\bibitem{AM2021} Y. Su {\it et al.}, Path-Dependent Thermal Metadevice beyond Janus Functionalities. {\it Adv.\ Mater.\/} {\bf 33}, 2003084 (2021).

\bibitem{WIL} A. Bejan, {\it Convection\ Heat\ Transfer\/} (John Wiley \& Sons, 2013).

\bibitem{PRL2018} D. Torrent, O. Poncelet, J.-C. Batsale, Nonreciprocal Thermal Material by Spatiotemporal Modulation. {\it Phys.\ Rev.\ Lett.\/} {\bf 120}, 125501 (2018).

\bibitem{SCI2019} Y. Li {\it et al.}, Anti–parity-time symmetry in diffusive systems. {\it Science\/} {\bf 364}, 170-173 (2019).

\bibitem{NM2019} Y. Li {\it et al.}, Thermal meta-device in analogue of zero-index photonics. {\it Nat.\ Mater.\/} {\bf 18}, 48–54 (2019).

\bibitem{NC2020} G. Xu {\it et al.}, Tunable analog thermal material. {\it Nat.\ Commun.\/} {\bf 11}, 6028 (2020).

\bibitem{AM2020} J. Li {\it et al.}, A Continuously Tunable Solid-Like Convective Thermal Metadevice on the Reciprocal Line. {\it Adv.\ Mater.\/} {\bf 32}, 2003823 (2020).

\bibitem{NP2022} G. Xu {\it et al.}, Diffusive topological transport in spatiotemporal thermal lattices. {\it Nat.\ Phys.\/} {\bf 18}, 450–456 (2022).

\bibitem{AIPA} S. Guenneau, D. Petiteau, M. Zerrad, C. Amra, T. Puvirajesinghe, Transformed Fourier and Fick equations for the control of heat and mass diffusion. {\it AIP\ Adv.\/} {\bf 5}, 053404 (2015).

\bibitem{PRE2018} G. Dai, J. Shang, J. Huang, Theory of transformation thermal convection for creeping flow in porous media: Cloaking, concentrating, and camouflage. {\it Phys.\ Rev.\ E\/} {\bf 97}, 022129 (2018).

\bibitem{ATE2021} B. Wang, T.-M. Shih, J. Huang, Transformation heat transfer and thermo-hydrodynamic cloaks for creeping flows: Manipulating heat fluxes and fluid flows simultaneously. {\it Appl.\ Therm.\ Eng.\/} {\bf 190}, 116726 (2021).

\bibitem{ATE2022} H. Wang, N.-Z. Yao, B. Wang, T.-M. Shih, X. Wang, Homogeneous Venturi-effect concentrators for creeping flows: Magnifying flow velocities and heat fluxes simultaneously. {\it Appl.\ Therm.\ Eng.\/} {\bf 206}, 118012 (2022).

\bibitem{PNAS2} Y. Liu, P. Tan, L. Xu, Kelvin–Helmholtz instability in an ultrathin air film causes drop splashing on smooth surfaces. {\it Proc.\ Natl.\ Acad.\ Sci.\ U.S.A.\/} {\bf 112}, 3280–3284 (2015).

\bibitem{PRL2019} H. Y. Lo {\it et al.}, Diffusion-Dominated Pinch-Off of Ultralow Surface Tension Fluids. {\it Phys.\ Rev.\ Lett.\/} {\bf 123}, 134501 (2019). 

\bibitem{PRAP2020} W.-S. Yeung, V.-P. Mai, R.-J. Yang, Cloaking: Controlling Thermal and Hydrodynamic Fields Simultaneously. {\it Phys.\ Rev.\ Appl.\/} {\bf 13}, 064030 (2020).

\bibitem{CPB} B. Wang, J. Huang, Hydrodynamic metamaterials for flow manipulation: Functions and prospects. {\it Chin.\ Phys.\ B\/} {\bf 31}, 098101 (2022).

\bibitem{PNAS3} J. Tong {\it et al.}, Bioinspired micro/nanomotor with visible light energy–dependent forward, reverse, reciprocating, and spinning schooling motion. {\it Proc.\ Natl.\ Acad.\ Sci.\ U.S.A.\/} {\bf 118}, e2104481118 (2021).

\bibitem{APL} E. M. Dede, T. Nomura, P. Schmalenberg, J. S. Lee, Heat flux cloaking, focusing, and reversal in ultra-thin composites considering conduction-convection effects. {\it Appl.\ Phys.\ Lett.\/} {\bf 103}, 063501 (2013).

\bibitem{SP} D.-P. Liu, P.-J. Chen, H.-H. Huang, Realization of a thermal cloak–concentrator using a metamaterial transformer. {\it Sci.\ Rep.\/} {\bf 8}, 2493 (2018).

\bibitem{PA} J. Li {\it et al.}, Doublet Thermal Metadevice. {\it Phys.\ Rev.\ Appl.\/} {\bf 11}, 044021 (2019).

\bibitem{Cry} C. Jiang, C. Fang, X. Shen, Multi-Physics Bi-Functional Intelligent Meta-Device Based on the Shape Memory Alloys. {\it Crystals\/} {\bf 9}, 438 (2019).

\bibitem{IJHMT} Y. Li {\it et al.}, Realization and analysis of an Intelligent flux transfer regulator by allocating thermal and DC electric fields. {\it Int.\ J.\ Heat\ Mass\ Transf.\/} {\bf 179}, 121677 (2021).

\bibitem{EPL} M. Lei, J. Wang, G. Dai, P. Tan, J. Huang, Temperature-dependent transformation multiphysics and ambient-adaptive multiphysical metamaterials. {\it Europhys.\ Lett.\/} {\bf 135}, 54003 (2021).

\bibitem{AM2022} J. Guo, G. Xu, D. Tian, Z. Qu, C.-W. Qiu, A Real-Time Self-Adaptive Thermal Metasurface. {\it Adv.\ Mater.\/} {\bf 34}, 2201093 (2022).

\end{thebibliography}
\end{document}